\documentclass{appolbnh}%
\usepackage{amssymb}
\usepackage{amsmath}
\usepackage{amsfonts}
\usepackage{graphicx}
\usepackage{slashed}%
\usepackage{multirow}
\begin{document}

\title{Scaling of the elastic proton-proton cross-section\thanks{Presented at ``Diffraction and Low-$x$ 2024'', Trabia (Palermo, Italy), September 8-14, 2024.}}
\author{\underline{Micha{\l} Prasza{\l}owicz}$^{(1)}$, Cristian Baldenegro$^{(1)}$, \\
Christophe Royon$^{(3)}$, Anna M. Sta{\'s}to$^{(4)}$
\address{
$^{(1)}$Institute of Theoretical Physics, 
Jagiellonian University,\\ S. {\L}ojasiewicza 11, 30-348 Krak{\'o}w, Poland,\\
$^{(2)}$Dipartimento di Fisica, Sapienza Universit{\' a} di Roma,\\ Piazzale Aldo Moro, 2, 00185 Rome, Italy,\\
$^{(3)}$Department of Physics and Astronomy,\\ 
The University of Kansas, Lawrence, KS 66045, USA, \\
$^{(4)}$Department of Physics, Penn State University, \\University Park, PA 16802, USA.}}

\maketitle

\begin{abstract}
We discuss scaling properties of the elastic $pp$ cross-section both
at the ISR and the LHC. We observe that the ratio of bump to dip
positions of the differential cross-section $d\sigma_{\rm el}/dt$ is constant 
over a wide energy range. We next study the consequences of this
property, including geometric scaling at the ISR and new scaling laws at the LHC.
 \end{abstract}


\section{Introduction}

In this note we recapitulate results on the scaling laws of the $pp$ cross-sections
(elastic, inelastic and total) discussed 
in Refs.~\cite{Baldenegro:2024vgg,Baldenegro:2022xrj}
and \cite{Royon:2022ych,Praszalowicz:2024pho}
(see also  \cite{Dremin:2012qd, Csorgo:2019ewn}).
We use data at two
energy ranges: ISR  \cite{Nagy:1978iw,Amaldi:1979kd}
covering $W=\sqrt{s} \simeq 20 \div 60~$GeV and  LHC TOTEM 
\cite{TOTEM:2011vxg}\nocite{TOTEM:2015oop,TOTEM:2017asr,TOTEM:2017sdy,TOTEM:2018psk}--\cite{TOTEM:2020zzr} 
of $W \simeq 3 \div 13~$TeV (the ATLAS measurements have focused 
only on the low $|t|$ region~\cite{ATLAS:2014vxr,ATLAS:2022mgx}). 

The energy behavior of  integrated $pp$ cross-sections at both energy ranges is very different.
At the ISR, elastic, inelastic and total cross-sections have almost the same
energy dependence \cite{Amaldi:1979kd}. This observation led
to the concept of geometric scaling \cite{DiasDeDeus:1973lde,Buras:1973km}.  This is not true at
the LHC~\cite{TOTEM:2017asr}.
In Ref.~\cite{Baldenegro:2024vgg} we  parametrized  $pp$,
cross-sections by the power laws shown
in Table \ref{tab:sigmas}. 

\renewcommand{\arraystretch}{1.2} \begin{table}[h]
\centering
\begin{tabular}
[c]{|c|c|c|c|c|}\hline
& elastic & inelastic & total & $\frac{\mathrm{elastic}}{\mathrm{inelastic}}$\\\hline
ISR & $W^{0.1142\pm0.0034}$ & $W^{0.1099\pm0.0012}$ & $W^{0.1098\pm0.0012}$ & $W^{0.0043\pm 0.0036}$ %
\\\hline
LHC & $W^{0.2279\pm0.0228}$ & $W^{0.1465\pm0.0133}$ & $W^{0.1729\pm 0.0163}$ & $W^{0.0814 \pm 0.0264}$ %
\\\hline
\end{tabular}
\caption{Energy dependence of the integrated cross-sections for the energies
$W=\sqrt{s}$ at the ISR \cite{Amaldi:1979kd}
and at the LHC \cite{Nemes:2019nvj}.}
\label{tab:sigmas}%
\end{table}\renewcommand{\arraystretch}{1.0}

Differential elastic $pp$ cross-sections
also reveal  significant differences, even though the general {\em dip-bump} structure is similar. 
The bump-to-dip cross-section ratio
\begin{equation}
\mathcal{R}_{\mathrm{bd}}(s)=\frac{  d\sigma_{\mathrm{el}} /d|t|_{\rm{b}}}
{ d\sigma_{\mathrm{el}} /d|t|_{\mathrm{d}}}%
\label{eq:Rbd}%
\end{equation}
saturates at the LHC 
 at  approximately 1.8, and  is rather strongly energy
dependent at the ISR (see e.g. Fig.~2 in Ref.~\cite{TOTEM:2020zzr}). 

However, even for differential cross-sections, there are some regularities that are common both to the ISR and the LHC. First,
the smallness of the real part of the forward elastic 
amplitude encoded in the
so called $\rho$ parameter. Second,
in Ref.~\cite{Baldenegro:2024vgg} we explored another regularity, namely  the ratio of  
bump-to-dip {\em positions} in $|t|$ at a given energy
\begin{equation}
\mathcal{T}_{\rm bd}(s)=|t_{\rm b}|/|t_{\rm d}|\, ,
\label{eq:Tbd}
\end{equation}
which is constant
at all energies from the ISR to the LHC and equal 
to $1.355\pm 0.011$ \cite{Baldenegro:2024vgg}, see Table~\ref{tab:dipbump}.
This suggests a 
scaling variable 
\begin{equation}
\tau= f(s) |t| \, ,
\label{eq:taudef}%
\end{equation}
where $f(s)$ is a universal function of energy.
Elastic differential cross-sections at different
energies, if plotted in terms of $\tau$, will have dips and bumps at exactly
the same values of $\tau_{\mathrm{d,b}}$.

\renewcommand{\arraystretch}{1.2} 
\begin{table}[h!]
\centering
\begin{tabular}[c]{|c|r|lc|cl|cc|}\hline
\multirow{2}{*}{~}& \multirow{2}{*}{$W$~} & \multicolumn{2}{c|}{dip} &\multicolumn{2}{c}{bump} & \multicolumn{2}{|c|}{ratios}\\
\cline{3-8}
&   & $\left\vert t \right\vert _{\mathrm{d}} $ & error
&$\left\vert t \right\vert _{\mathrm{b}} $ & ~error
&$t_{\mathrm{b}}/t_{\mathrm{d}}$ & error
\\\hline
\multirow{4}{*}{\rotatebox{90}{LHC [TeV]}} 
&  13.00 & 0.471 & $^{+0.002}_{-0.003}$ & ~0.6377 & $^{+0.0006}_{-0.0006}$ & 1.355 & $^{+0.008}_{-0.005}$\\
&  8.00 & 0.525 & $^{+0.002}_{-0.004}$ & 0.700 & $^{+0.010}_{-0.008}$ & 1.335 &$^{+0.021}_{-0.016}$\\
&  7.00 & 0.542 & $_{-0.013}^{+0.012}$ & 0.702 & $^{+0.034}_{-0.034}$ & 1.296 &$^{+0.069}_{-0.069}$\\
&  2.76 & 0.616 & $^{+0.001}_{-0.002}$ & 0.800 & $^{+0.127}_{-0.127}$ & 1.298 &$^{+0.206}_{-0.206}$ \\
\hline
\multirow{5}{*}{\rotatebox{90}{ISR [GeV]}} 
&  62.50 & 1.350 & $^{+0.011}_{-0.011}$ & 1.826 & $^{+0.016}_{-0.039}$ & 1.353 & $^{+0.016}_{-0.029}$\\
&  52.81 & 1.369 & $^{+0.006}_{-0.006}$ & 1.851 & $^{+0.014}_{-0.018}$ & 1.352 & $^{+0.012}_{-0.014}$\\
&  44.64 & 1.388 & $^{+0.003}_{-0.007}$ & 1.871 & $^{+0.031}_{-0.015}$ & 1.348 & $^{+0.023}_{-0.011}$\\
& 30.54 & 1.434 & $_{-0.004}^{+0.001}$ & 1.957 & $_{-0.028}^{+0.013}$ & 1.365 &$_{-0.020}^{+0.010}$\\
&  23.46 & 1.450 & $_{-0.004}^{+0.005}$ & 1.973 & $_{-0.018}^{+0.011}$ & 1.361 & $_{-0.013}^{+0.009}$\\
\hline
\end{tabular}
\caption{ Positions  of bumps  
obtained by fitting a parabola, and position of dips obtained by fitting a parabola (LHC) or a third order polynomial (ISR),
Ref.~\cite{Baldenegro:2024vgg}.
}%
\label{tab:dipbump}%
\end{table}
\renewcommand{\arraystretch}{1.0} 

\section{Geometric scaling at the ISR}

\label{sec:GS@ISR}

Unitarity constraints allow to write scattering cross sections in the impact
parameter space,
\begin{align}
\sigma_{\mathrm{el}}  & =%
{\displaystyle\int}
d^{2}\boldsymbol{b}\,\left\vert 1-e^{-\Omega(s,b)+i\chi(s,b)}\right\vert
^{2},\nonumber\\
\sigma_{\mathrm{tot}}  & =2%
{\displaystyle\int}
d^{2}\boldsymbol{b}\,\operatorname{Re}\left[  1-e^{-\Omega(s,b)+i\chi
(s,b)}\right]  ,\nonumber\\
\sigma_{\mathrm{inel}}  & =%
{\displaystyle\int}
d^{2}\boldsymbol{b}\,\left[  1-\left\vert e^{-\Omega(s,b)}\right\vert
^{2}\right] \, , 
  \label{eq:sigmas}%
\end{align}
in terms of the opacity  $\Omega(s,b)$ and the phase $\chi(s,b)$, 
which is responsible for the nonzero $\rho$ parameter  \cite{Barone:2002cv,Levin:1998pk}.
 However, since the $\rho$ parameter is very small, we can neglect $\chi(s,b)$
in the first approximation.

Geometric scaling (GS) is a hypothesis \cite{DiasDeDeus:1973lde}  that%
\begin{equation}
\Omega(s,b)=\Omega\left(  b/R(s)\right)\, , \label{eq:GSdef}%
\end{equation}
where $R(s)$ is the interaction radius \cite{DiasDeDeus:1973lde} increasing
with energy. Changing the integration variable in (\ref{eq:sigmas}) $\boldsymbol{b}\rightarrow \boldsymbol{B}=\boldsymbol{b}/R(s)$
leads to
\begin{equation}
 \int d^b\boldsymbol{b} \cdots=
R^2(s) \int d^2\boldsymbol{B} \cdots \, .
\label{eq:sigmas1}
\end{equation}
Here the integral over $d^{2}\boldsymbol{B}$ is an energy-independent constant. Therefore, the 
integrated $pp$  cross-sections should scale with energy in the same way. As seen from Table~\ref{tab:sigmas}, this is indeed the case.

\begin{figure}[h]
\centering
\includegraphics[height=5.5cm]{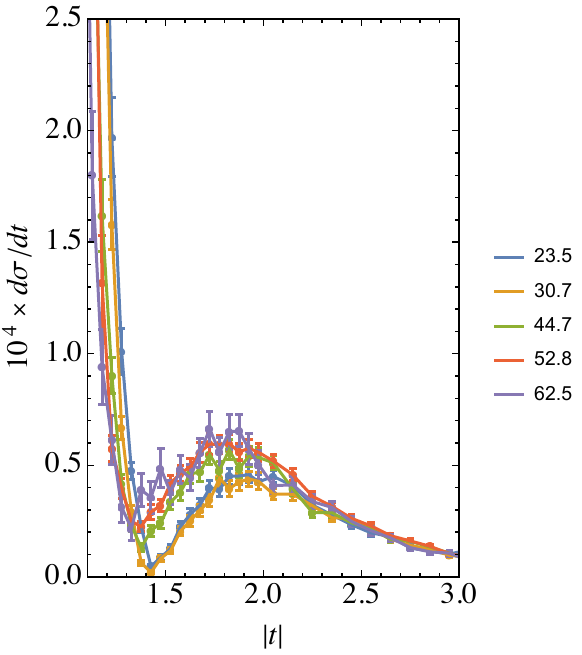}
~\includegraphics[height=5.5cm]{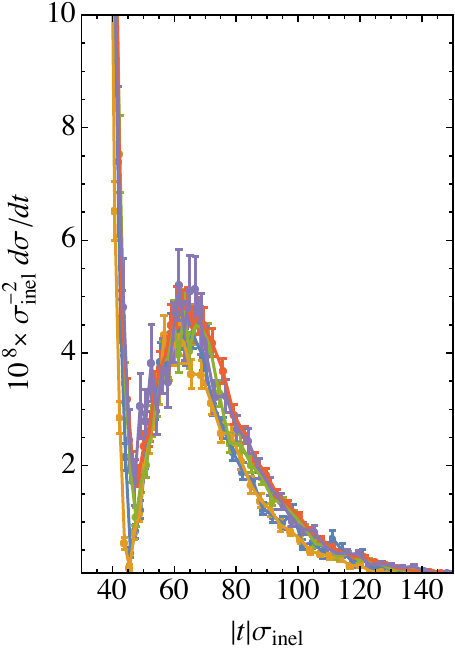}\caption{Elastic pp cross-section
$d\sigma_{\mathrm{el}}/dt$~[mb/GeV$^{2}$] 
at the ISR 
multiplied by $10^4$
-- left,  multiplied by $10^8$ 
and scaled according to
Eqs.~(\ref{eq:BDdDscaling},\ref{eq:tau}) -- right.
}
\label{fig:isr}%
\end{figure}

Ref. \cite{Buras:1973km} analyzed the consequences of GS for the differential
rather than integrated
$pp$ elastic cross-section assuming the following scaling variable
\begin{equation}
\tau=\sigma_{\mathrm{inel}}(s)\,|t|\,=R^{2}(s)|t|\times\mathrm{const.}%
\label{eq:tau}%
\end{equation}
One can show \cite{Buras:1973km} that in this case the function $\Phi(\tau)$ defined as
\begin{equation}
\Phi(\tau)=\frac{1}{\sigma_{\mathrm{inel}}^{2}(s)}\frac{d\sigma_{\mathrm{el}}}%
{d|t|}(s,t) \label{eq:BDdDscaling}%
\end{equation}
should not depend on energy. In Fig.~\ref{fig:isr} we plot the ISR data 
before and after scaling. We see that
the  cross-sections
overlap after scaling, except for the dip region. In Ref.~\cite{Baldenegro:2024vgg}
we have quantified the quality of this overlap by plotting  ratios of the scaled cross-sections in terms of $\tau$.

\section{The LHC scaling}
\label{sec:GS@LHC}

As can be seen from Table~\ref{tab:sigmas} the energy behavior of the $pp$ cross-sections
at the LHC
is not universal and therefore GS does not hold, although the scaling variable
(\ref{eq:taudef}) still superimposes positions of dips and bumps 
(but not the values of the cross-sections).
However, 
at the LHC 
ratios
$\mathcal{R}_{\mathrm{bd}}$ (\ref{eq:Rbd}) are
(almost) energy independent, which suggests a new universal behavior
\begin{equation}
\frac{d\sigma_{\mathrm{el}}}{dt}(t_{\mathrm{d}})=g(s)\, \mathrm{const}%
_{\mathrm{d}},~~\frac{d\sigma_{\mathrm{el}}}{dt}(t_{\mathrm{b}})=g(s) \,
\mathrm{const}_{\mathrm{b}} \, .
\label{eq:xsecscaling}%
\end{equation}
Therefore, we expect that dips and bumps at the LHC
will overlap when the differential cross-section will be scaled by two functions $f(s)$ and $g(s)$ rather
than by one function as was the case at the ISR.\footnote{Strictly speaking the cross-section is scaled by $g$ and the scaling
variable by $f$.} 

\begin{figure}[h]
\centering
\includegraphics[height=5.5cm]{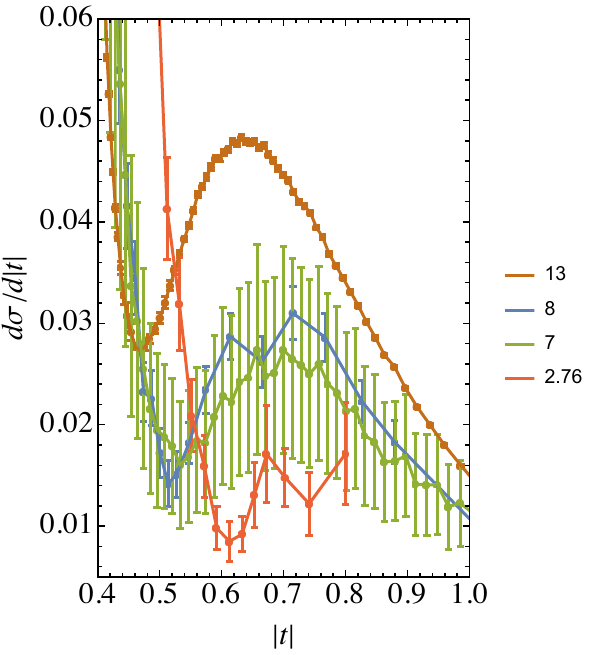}
~\includegraphics[height=5.5cm]{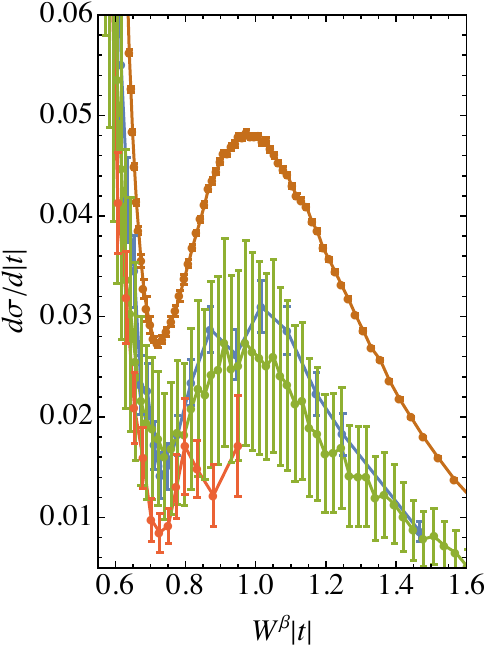}
\caption{Elastic $pp$ cross-section
$d\sigma_{\mathrm{el}}/dt$~[mb/GeV$^{2}$] at the LHC energies  in terms of $|t|$~[GeV$^{2}$] -- left, and in terms of the
scaling variable $W^{\protect\beta}\,|t|$ -- right. We 
can see that dip and
bumps are aligned after scaling.}%
\label{fig:lhc1}%
\end{figure}

Fitting dip and bump positions 
with a power law, i.e. $f(W)=B\, W^{-\beta}$ in Eq.~(\ref{eq:taudef}) 
 one obtains
\begin{align}
t_{\mathrm{dip}}(W) &=(0.732\pm 0.003)\times(W/(1~\mathrm{TeV}))^{-0.1686\pm 0.0027}\, , \notag \\
t_{\mathrm{bump}}(W) & =1.355 \times t_{\mathrm{dip}}(W) \, .
\label{eq:dipbumpfits}%
\end{align}
Now, if we plot $d\sigma_{\mathrm{el}}/d|t|$ in
terms of the variable $\tau=(W/(1~\mathrm{TeV}))^{0.1686}\, |t|$, the dip and bump positions
are aligned, as seen in Fig.~\ref{fig:lhc1}. 

\begin{figure}[h]
\centering
\includegraphics[height=5.2cm]{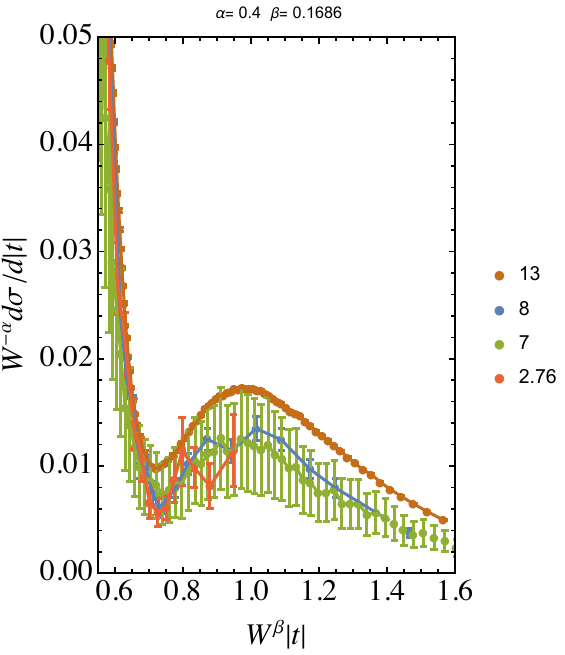}
\includegraphics[height=5.2cm]{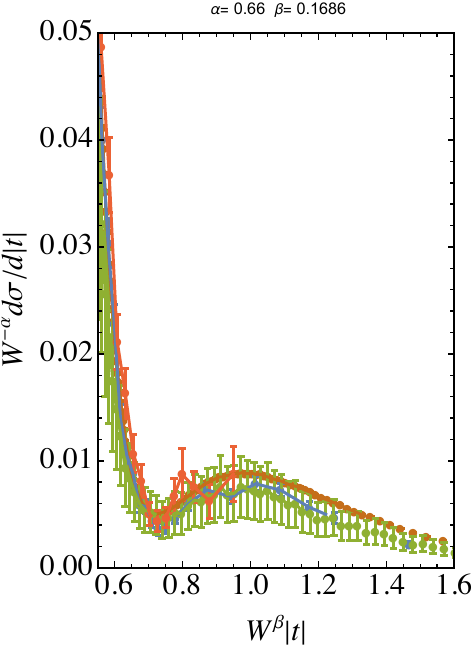}
\includegraphics[height=5.2cm]{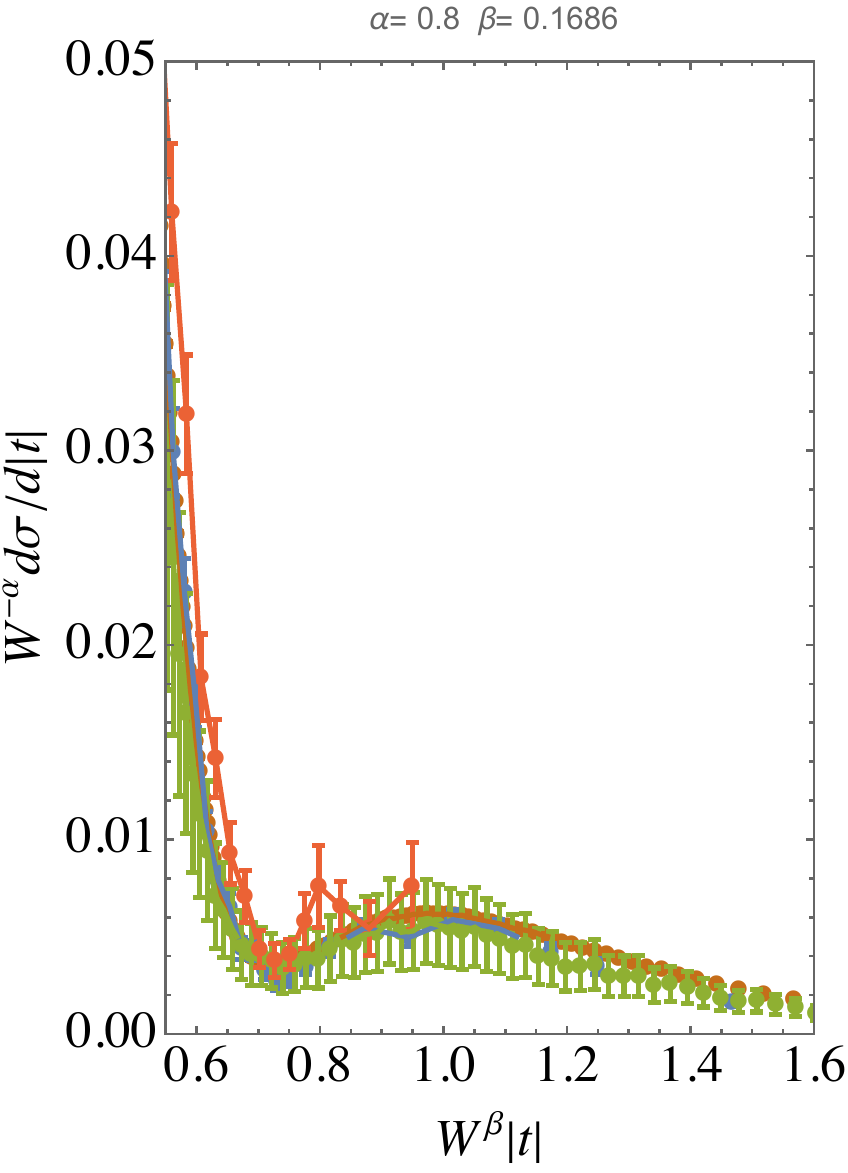}
\caption{Scaled elastic $pp$
cross-section $d\sigma_{\mathrm{el}}/dt$~[mb/GeV$^{2}$] at the LHC energies
 in terms of the scaling variable
$W^{\beta}|t|$ for $\beta=0.1686$ and for $\alpha=0.4$ (left), $\alpha=0.66$ (middle) and
$\alpha=0.8$ (right).}%
\label{fig:lhcfull}%
\end{figure}

Next, we aim at superimposing all curves in the right panel of
Fig.~\ref{fig:lhc1} shifting them vertically by an energy dependent factor.
To this end we try a simple transformation
(\ref{eq:xsecscaling}),
\begin{equation}
\frac{d\sigma_{\mathrm{el}} }{d|t|}(\tau)\rightarrow\left(  \frac
{W}{1\mathrm{TeV}}\right)  ^{-\alpha}\frac{d\sigma_{\mathrm{el}} }{d|t|}%
(\tau) \, .
\label{eq:salpha}%
\end{equation}
The results are shown in Fig.~\ref{fig:lhcfull} 
for three values of $\alpha$ for fixed $\beta=0.1686$. In the three panels of Fig.~\ref{fig:lhcfull} 
elastic cross-sections overlap 
or nearly overlap in comparison with the right panel of Fig.~\ref{fig:lhc1}, indicating scaling. 
In Ref.~\cite{Baldenegro:2024vgg} we  estimated  the best value to be $\alpha=0.66$ with large systematic error.
Obviously, global fits may result in different values of $\alpha$ and $\beta$. 

\section{Other scaling laws}\label{sec:otherScalingLaws}
In Ref.~\cite{Baldenegro:2022xrj} another scaling law for the LHC data was
 proposed. There $\Phi(\tilde{\tau})$ (\ref{eq:BDdDscaling}) depended on
 the scaling variable defined as
\begin{equation}
\tilde{\tau}=s^{a} t^{b}. \label{eq:BRStau}%
\end{equation}
Best values of $a$ and $b$ were obtained by minimizing the so called {\it quality factor}
(QF) 
\cite{Gelis:2006bs,Beuf:2008mf,Marquet:2006jb}
with the result $\alpha\simeq0.61$ 
and $a \simeq0.065$, $b \simeq0.72$.

To relate this scaling  to
the scaling  of Sec.~\ref{sec:GS@LHC} let's note that if
(\ref{eq:BRStau}) should align the dips (and
bumps) at all LHC energies, the scaled dip (bump) positions
\begin{equation}
\tilde{\tau}_{\mathrm{d}}=s^{a} t_{\mathrm{d}}^{b}=s^{a-b\, \beta/2}B_{\mathrm{dip}%
}^{b}\label{eq:BRStau1}%
\end{equation}
should be energy independent. Here we used
(\ref{eq:taudef}) with $f(s)=B\, s^{-\beta/2}$. From the energy independence of
(\ref{eq:BRStau1}) we obtain
\begin{equation}
a-b\, \beta/2 =0.\label{eq:zeroabbeta}%
\end{equation}
Substituting $b$ from \cite{Baldenegro:2022xrj} and  $\beta= 0.1686$ we find
$a=b\, \beta/2 =0.061\pm 0.001$ as compared to $a=0.065$ from Ref.~\cite{Baldenegro:2022xrj}.
Furthermore, our result for the power $\alpha=0.66$, which is poorly constrained by the data,
is again in qualitative agreement with \cite{Baldenegro:2022xrj}, where
$\alpha=0.61$.
Obviously Eq.~(\ref{eq:zeroabbeta}) implies the whole family of scaling laws.
In a less restricted setup also the parameter $\beta$ should be determined from global fits.

\section{Summary and conclusions}\label{sec:summary}

We explored the property that
 ($t_{\mathrm{b}%
}/t_{\mathrm{d}}$) 
is constant over three orders of magnitude in energy. 
This behavior suggests 
a universal behavior of the elastic scattering cross-sections as a function of a scaling variable $\tau=f(s)|t|$.

Such transformation was sufficient at the ISR, but at the LHC one had to modify the values of the cross-sections
by another scaling function $g(s)$. This difference may be related to the saturation \cite{Baldenegro:2022xrj}.
For the consequences of the above scaling for the phenomenological parameterizations of the scattering
amplitude, see Ref.~\cite{Baldenegro:2024vgg}.

\section*{Acknowledgments}

MP thanks the organizers for a fruitful and stimulating workshop. 
AMS is supported by the U.S. Department of Energy grant No. DE-SC-0002145 and within the framework of the
Saturated Glue (SURGE) Topical Theory Collaboration. CB is supported by the European Research Council consolidator grant no. 101002207.

\end{document}